# Experimental investigations of psychoacoustic characteristics of household vacuum cleaners

Sanjay Kumar[(a)], Wong Sze Wing, Teng Mingbang, and Heow Pueh Lee[(b)]

Department of Mechanical Engineering, National University of Singapore, 9 Engineering Drive 1, Singapore 117575, Singapore

Corresponding author's email: a) mpesanj@nus.edu.sg, b) mpeleehp@nus.edu.sg

## Abstract

Vacuum cleaners are one of the most widely used household appliances associated with unpleasant noises. Previous studies have indicated the severity of vacuum cleaner noise and its impact on the users nearby. The quantified measurements of the generated noise standalone are not sufficient for the selection or designing of vacuum cleaners. The human perception must also be included for a better assessment of the quality of sound. A hybrid approach known as psychoacoustics, which comprises subjective and objective evaluations of sounds, is widely used in recent times. This paper focuses on the experimental assessment of psychoacoustical matrices for household vacuum cleaners. Three vacuum cleaners with different specifications have been selected as test candidates, and their sound qualities have been analyzed. Besides, the annoyance index has been assessed for these vacuum cleaners.

## 1. Introduction

In almost every house, several electromechanical home appliances are used, such as fans, hair driers, grinders, juicers, microwave ovens, and vacuum cleaners. Among these, the vacuum cleaner is one of the most frequently used home appliances which generates unpleasant noise during the operations. Many people ignore one crucial aspect, namely the noise level when buying a vacuum cleaner. The noise is mostly emitted from the built-in suction units of the vacuum cleaners, exhaust fan, the airflow, and the surface vibrations during the operation [1]. The suction unit generally consists of a driving electric motor and a centrifugal blower. In some equipment, a vanned diffuser is installed in the hose to achieve high-pressure rise during the operation. A detailed description of noise generation source(s) of typical vacuum cleaners can be found in various publications [2-6]. The sound pressure level of working vacuum cleaners is varied from 65 dB(A) to 90 dB(A). A long-term exposure to excessive levels of noise (≥80 dBA) can lead to several adverse human health issues such as, stress, fatigue, psychological disorders, hearing loss, high blood pressure, coronary heart disease, sleeping disorder, hypertension, obesity, cognitive impairment in children, and diabetic type I and II, hence making it imperative to reduce the noise levels [7-9]. The risk is typically low for normal usage but extended usage for professional cleaners will be exposed to higher risk.

Recently, the European Union released eco-design regulation guidelines related to energy consumptions and noise levels for vacuum cleaners. The maximum motor size and the peak noise level for any vacuum cleaners in operation are limited to 900 W and 80 dB(A), respectively [10, 11]. However, manufacturers have reported their concerns for the full implementation of those directives, because the noise levels are sometimes psychologically misinterpreted by the users as the cleaning capacity of the vacuum cleaners. Symanczyk [12] pointed out the paradox with vacuum cleaner sounds: "you can make them very silent, but then they will not be perceived as very powerful". Some vacuum cleaners which fail to meet EU guidelines are still available in the market for sale [13]. In recent times, the European Union has released strict directives to the vacuum cleaner manufacturers for further reduction in operating noise.

Owing to these international regulations, a comprehensive assessment of sound qualities for the commercial vacuum cleaners is necessary. In this regard, several research studies have been reported [14]. Researchers started from the measurement of sound pressure levels of the generated noise and assessed the performance of the equipment as per international regulations ($LA_{eq}$ < 80 dBA of the 8-hour daily occupational noise exposure). However, the sound pressure level data stand-alone is not sufficient for any conclusion made for the noise source because human perception towards the surrounding noise is different from the quantified sound levels. The human perception for the sound considers subjective and objective elements and may vary from person-to-person. Apart from the sound intensity of the noise, other subjective parameters like environmental conditions, psychological factors (pleasantness), psychoacoustical parameters (loudness, sharpness, roughness, fluctuation strength, and tonality) need to be incorporated in the sound quality assessments [15-19]. In recent years, several psychoacoustical studies have been reported for household appliances. Ih et al. [20] experimentally investigated the product sound quality of the vacuum cleaner. First, the Taguchi orthogonal array method, an experimental design technique was utilized to assess the frequency range that described the sound quality of the vacuum cleaners. Then, a psychoacoustic model 'annoyance index' was developed by using linear regression analysis. The developed model was further validated by the artificial intelligence-based technique, artificial neural network (ANN). Takada et al. [21] conducted a psychoacoustic assessment for the economic evaluation of the sound quality index for vacuum cleaners and hair dryers. Conjoint analysis was used to evaluate the buying willingness of these items based on generated noise. It was reported that the person's purchasing willingness increased for the equipment with the lower values of sound pressure level and sharpness. Rukat et al. [6] investigated the effect of operating conditions on vacuum cleaner noise. They reported that the vacuum cleaner emitted a higher level of noise while working on a flat surface like terracotta as compared to working on the carpet. Their conclusion was based on the measurement of A-weighted sound pressure levels of the tested vacuum cleaners. Moravec et al. [22] developed a sound quality index for automatic washing machine noise. Their psychoacoustic model was based on the correlation between subjective and objective sound quality assessment. Novakovic et al. [23] experimentally validated the previously correlation models proposed by Lipar et al. [16] and Di et al. [24]. Their findings suggested that both models were appropriate for quick assessment of sound quality index. Ma et al. [25] investigated the influence of the dental equipment noise on the perceptions and behaviors of dental professionals. More recently, Murovec et al. [26] conducted psychoacoustic analysis for cavitation detection in centrifugal pumps. These reported works confirm the broad applications of psychoacoustic assessment of noise.

In this work, we have performed an experimental assessment on the psychoacoustic characteristics of commercially available household vacuum cleaners. Three dry-type vacuum cleaners from different manufacturers are selected, and their acoustical performances have been investigated. In the study, first, we measured the noise levels generated from the selected vacuum cleaners, and the recorded data is used in the evaluation of psychoacoustic parameters, namely, loudness, sharpness, roughness, and fluctuation strength. Furthermore, the Annoyance index, a psychological attribute of psychoacoustics, has been evaluated for these vacuum cleaners. The annoyance index was based on the previously developed annoyance index equation by Altmsoy et al. [27] for dry type vacuum cleaners.

## 2. Psychoacoustic parameters

### 2.1 Loudness

Loudness is one of the most important subjective parameters of the psychoacoustic analysis. The loudness is slightly different from the sound amplitude. The sound amplitude is a measured sound intensity in decibel value whereas loudness is the psychological aspects of the measured sound pressure level. The former is quantified by the microphone perceived sound, and the later involves the human perception of the sound. It is interesting to note that the human ear can perceive sounds of different frequencies at different sound pressure levels as equally loud. This is due to the differing sensitivity of the human ear within the hearing domain. In 1933, Fletcher-Munson developed the curves of equal loudness to accurately define perceived loudness across all frequency spectrums within the human hearing domain. The calculation of loudness (L) is made by several approved standards such as ISO532B and DIN 45631 which are also available in various sound processing software. Loudness level is expressed in phons (P) or sones (S). The $phon$ is a unit of loudness that represents equal loudness to a 1000Hz pure tone, whereas sones is defined as the loudness perceived by typical listeners when confronted with a 1000 Hz tone at a sound pressure level of 40 phons. The loudness level in sones can be calculated from the relation $s = 2^{(P-40)/10}$.[28]

### 2.2 Sharpness

Sharpness delineates the human sensation caused by high frequency components of the noise. The units of sharpness (SH) is $acum$. A narrow band noise at 1 kHz with a bandwidth of < 150 Hz and a sound level of 60 dB is defined as 1 $acum$. The variability of sharpness depends upon the specific loudness distribution of the sound. It is calculated as a weighted area of loudness from the relation $SH = 0.11 \frac{\int_0^{24\ Bark} N' g(z) z dz}{\int_0^{24\ Bark} N' dz}$, where $N'$ is the specific loudness that exhibits the loudness distribution across the critical-bands, $g(z) = \begin{cases} 1; & for\ f < 3kHz \\ 4; for\ 3\ kHz < f < 20\ kHz \end{cases}$ is the weighting function, and $z$ is the critical-band rate defined by Zwicker.[17]

### 2.3 Fluctuation Strength and Roughness

When multiple tones are modulated or combined to form a single sound, the sound level rises and falls over time. These sound level fluctuations arise from the constructive and destructive interferences of the tones for different frequency. Examples of such sounds could be the siren from an ambulance or the rumbling of a car engine. The amount of sound modulation determines the sensation that is perceived by the human ear. This sensation can be modeled by two analytical parameters, namely, fluctuation strength and roughness [14]. These psychoacoustic matrices quantify the amount of sound modulation based on the following aspects: modulation frequency – the number of rises and falls in the sound per sound, and modulation Level – the perceived magnitude level change over time.

The fluctuation Strength describes with 20 modulations per second or less (between 0.5 and 20 Hz), and the roughness is perceived as the sensation of rapidly modulated sound or vibration in the modulation range between 20 and 300 times per second (20 – 300 Hz). In terms of the human hearing domain, the ear can pick up modulations below 20 Hz, and anything beyond it, perceived as a stationary and rough tone. These two metrics can invoke individual sensory perceptions. For instance, a tone with high fluctuation strength is regarded as an alert (fire alarms, sirens) while high roughness has been used in automotive industries to accentuate the sportiness of the vehicle. The units to describe fluctuation strength (F) and roughness (R) is $vacil$ and $asper$, respectively. 1 $vacil$ is defined as a tone with sound pressure level of 60 dB at 1 kHz modulated by 100 % at modulation frequency of 4 Hz. While, 1 $asper$ represents the roughness produced by a pure tone at 1000 Hz and with SPL of 60 dB which is modulated by 100 % modulation depth and modulation frequency ($f_{mod}$) of 70 Hz. The roughness of sound can be estimated from the equation, $R = cal \times \int_0^{24\,Bark} f_{mod}\,\Delta L\,dz$, and the fluctuation strength is determined from the equation, $F = \frac{0.008 \times \int_0^{24\,Bark} f_{mod}\,\Delta L\,dz}{\left(\frac{f_{mod}}{4\,Hz}\right)+\left(\frac{4\,Hz}{f_{mod}}\right)}$, where $\Delta L$ is the perceived masking depth, $cal$ is the calibration factor [17].

### 2.4 Annoyance Index

The annoyance index is one of the most important attributes for the psychoacoustical assessment. It describes the human sentiments towards the incoming sound waves. In the annoyance index analysis, the combination of the psychological parameters and the psychoacoustical matrices are included. The physical parameters can be collected through jury testing in which data are collected from the persons directly. In the process, the recorded sound generated from appliances such as vacuum cleaners are played for a short period in front of listeners, and some psychological questions are asked, and their responses are rated on a point scale. The results of the subjective judgments are used to generate a value of perceived annoyance factor for each vacuum cleaner. The collected data are then correlated using regression analysis with the psychoacoustic parameters measured from the same vacuum cleaners. The study gives rise to an Annoyance index (AI), which can be used to determine the sound quality of equipment. Altmsoy et al. [27] developed a general annoyance index equation for assessment of the sound qualities of vacuum cleaners. The annoyance index (AI) for a vacuum cleaner can be expressed as a linear combination of psychoacoustics metrics, $AI = 0.1(L + SH + 15R + 5F)$, where $L$ is loudness, $SH$ is sharpness, $R$ is roughness, and $F$ is fluctuation strength [27].

## 3. Experimental design and methods

### 3.1 Test Environment

The experiment was carried out in a room of the Vibration/Dynamics Laboratory located in the National University of Singapore. The room was 5.9 m x 8.3 m in size with 2 doors and several cabinets surrounding 3 sides of the walls. There were 4 tables in the room and various equipment was placed onto them. **Figure 1**(a) shows the floor plan and the photo of the test room. The room was selected to perform the experiment since it had similar size and obstructions to a living room, which was a common operating environment for a vacuum cleaner.

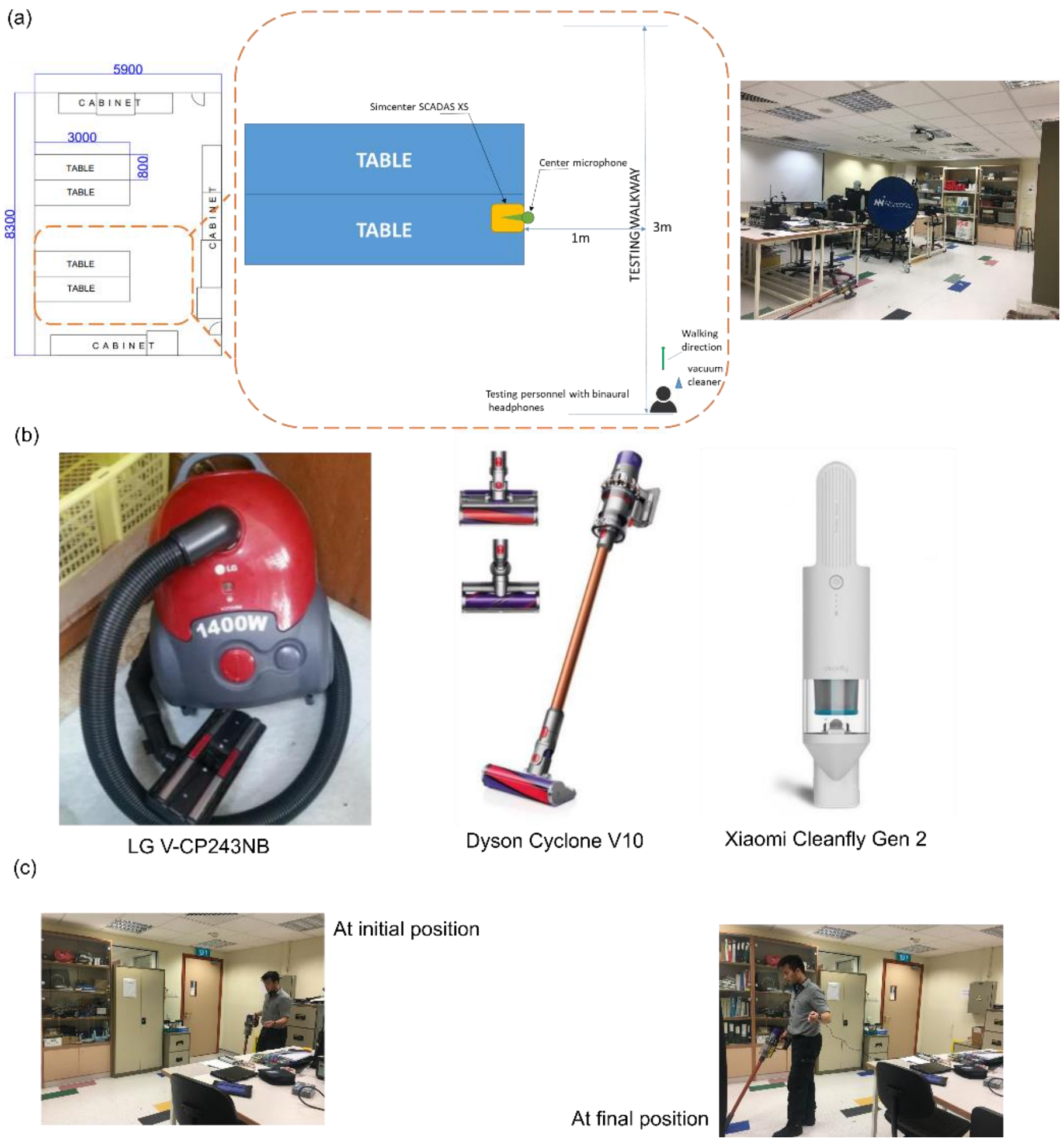

**Figure 1. (a)** Schematics of experimental setup plan. **(b)** Photographs of the three different vacuum cleaners. LG V-CP243NB (model type 1), Dyson Cyclone V10 (model type 2), and Xiaomi Cleanfly Gen2 (model type 3). **(c)** Photographs of the testing room displaying the vacuum cleaner positions.

## 3.2 Test appliances

Three different makers of dry-type vacuum cleaners were selected for the psychoacoustic studies. The manufacturers launched the selected vacuum cleaners in three different periods. The LG V-CP243NB is the oldest model, while the Dyson Cyclone V10 and Xiaomi models are recently launched. **Figure 1**(b) shows photographs of the selected models. LG V-CP243NB is a traditional vacuum cleaner that includes a suction unit on the sled and a flexible hose for transporting the dust. The power consumption is 1400 W with a single power setting. Dyson Cyclone V10 is an upright vacuum cleaner that is well known for its fourteen patented concentric cyclones feature to allow efficient and accelerated airflows for capturing the microscopic particles as small as 0.3 microns. Their digital motor consumes 525 W to generate 125,000 rpm motor speed for 151 air watts suction power. It comes with three power settings to suit different tasks, and the running time is up to 60 mins. The overall weight of the device is 2.5kg. Xiaomi Cleanfly Gen 2 is a portable wireless handheld vacuum cleaner. It consists of a brushless DC motor with a maximum rotation speed of 100,000 rpm with a total power consumption of 120W. Xiaomi Cleanfly comes with two-speed settings for different applications. It is capable of developing a maximum suction pressure of 16.8 kPa and comes with a compact design with an overall weight of 560g. The product specifications of these three vacuum cleaners are listed in **Table 1**.

**Table 1.** Product specifications of the used vacuum cleaners.

| Specifications | LG V-CP243NB | Dyson Cyclone V10 | Xiaomi Cleanfly Gen2 |
|---|---|---|---|
| Annotation used | Model type 1 | Model type 2 | Model type 3 |
| Power rating (Watt) | 1400 | 525 | 120 |
| Suction pressure | N.A.* | 151 air Watt | 16.8 kPa |
| Motor speed (R.P.M.) | N.A.* | 1,25,000 | 1,00,000 |
| Overall weight (kg) | 3.5 | 2.5 | 0.56 |

*N.A. = Not available.

### 3.3 Test equipment and signal processing

The ambient and vacuum cleaner noise was recorded by a handheld data acquisition system (The Simcenter SCADAS XS-Siemens®) [29]. It is capable of acquiring dynamic data simultaneously at 50,000 samples per second on up to 12 dynamic channels. The portable hardware was connected with a binaural headset and a central microphone. This headset is equipped with two microphones positioned at both ears to make binaural recordings. A ¼" pressure field central microphone (PCB) was also connected with the data acquisition system. The data recorded from the binaural headset reflected the noise influence to the vacuum cleaner user whilst the regular microphone was used to capture the noise effect from the vacuum cleaner to the surrounding occupants. The SCADAS system was remotely controlled by the Simcenter Testlab Scope App software installed on an Android-based Samsung tablet. The recorded data was post-processed in the LMS test lab software (Siemens®), and several psychoacoustic parametric values were extracted.

### 3.4 Noise measurement methods

For noise measurements of the vacuum cleaners, the walking test was preferred over the static test. Because, in the walking test methods, suction noise, and the noise generated from the interaction between the vacuum cleaner and the floor, are both considered. While in the static test, only suction noise is measured. In the test procedure, SCADAS XS and a central microphone were placed on the table (**Figure 1**a). The central microphone was used to investigate the noise annoyance towards the surrounding occupants. Also, a set of binaural headphones was placed on the test personnel's shoulder for recreation of the sound disturbance perceived by the user. The central microphone was kept at a distance of 1 m away from the walkway, and the walking distance was limited to 3 m as well. The microphones were calibrated by using a standard sound calibrator. The test personnel operated the vacuum cleaners while walking for a linear distance of 3 m, and the generated noise was recorded. Prior to this, background noise was first recorded in the same experimental condition, and the results were used for comparison with the acoustic characteristics for a vacuum cleaner noise. **Table 2** enlists the annotations used to represent the microphones and motor speed settings.

For equivalent sound pressure level calculations, 32 readings of sound pressure level (SPL) were recorded at an interval of 30 seconds in a period of 10 minutes, and their minimum and maximum SPL were extracted. From these data, the equivalent noise level calculated by using the following expressions, $LA_{eq} = 10\log\left(\sum_{i=1}^{n}\frac{\alpha i\times 10\times L}{10}\right)$, where $n$ is the number of observations, $\alpha$ is the fraction of time for which SPL persists, $i$ is the time interval, $L$ is the sound intensity. In this study, we have taken $\alpha = 0.9$ for calculation of equivalent sound pressure levels.

**Table 2.** Annotations used in the study.

| **Annotations** | **Descriptions** |
|---|---|
| C1 | Centre Microphone |
| C2 | Right Ear Microphone (from the headphones) |
| C3 | Left Ear Microphone (from the headphones) |
| S1 | Minimum motor speed setting of vacuum cleaners |
| S2 | Maximum motor speed setting of vacuum cleaners |

## 4. Results and discussion

### 4.1 Acoustical performances

**Figure 2** (a-c) depicts the measured A-weighted sound levels spectra (one-third octave) for the three different vacuum cleaners; model type 1, model type 2, and model type 3, respectively. All measurements were performed at the same venue, under similar environmental conditions, and by the same person. Each experiment has repeated ten times, and their average values have considered in the study. As shown in **Figure 2**(a), the noise level of the model type 1 vacuum cleaner is maximum in the mid-range frequencies (500-2000 Hz). The Equivalent pressure level $LA_{eq}$ for C1, C2, and C3 are 77.4 dBA, 90.4 dBA, and 80.6 dBA, respectively. In the case of model type 2, the maximum noise level is observed in the mid-frequency range at both speed settings. Also, the equivalent pressure level is measured to be 69.8 dBA, 86.2 dBA, and 70.7 dBA for C1, C2, and C3 in the minimum setting while the $LA_{eq}$ is 75.4 dBA, 94.8 dBA and 76.3 dBA for C1, C2, and C3 in the maximum setting. **Figure 2**(c) shows the sound pressure level spectrum obtained from the model type 3 vacuum cleaner. It is observed that the maximum sound level values lie in mid-range frequencies in both speed settings S1 and S2. The $LA_{eq}$ is 64.1 dBA, 79.6 dBA, and 70.4 dBA for C1, C2, and C3 in the minimum setting while the $LA_{eq}$ is 71.2 dBA, 90.0 dBA, and 79.2 dBA for C1, C2 and C3 in the maximum setting.

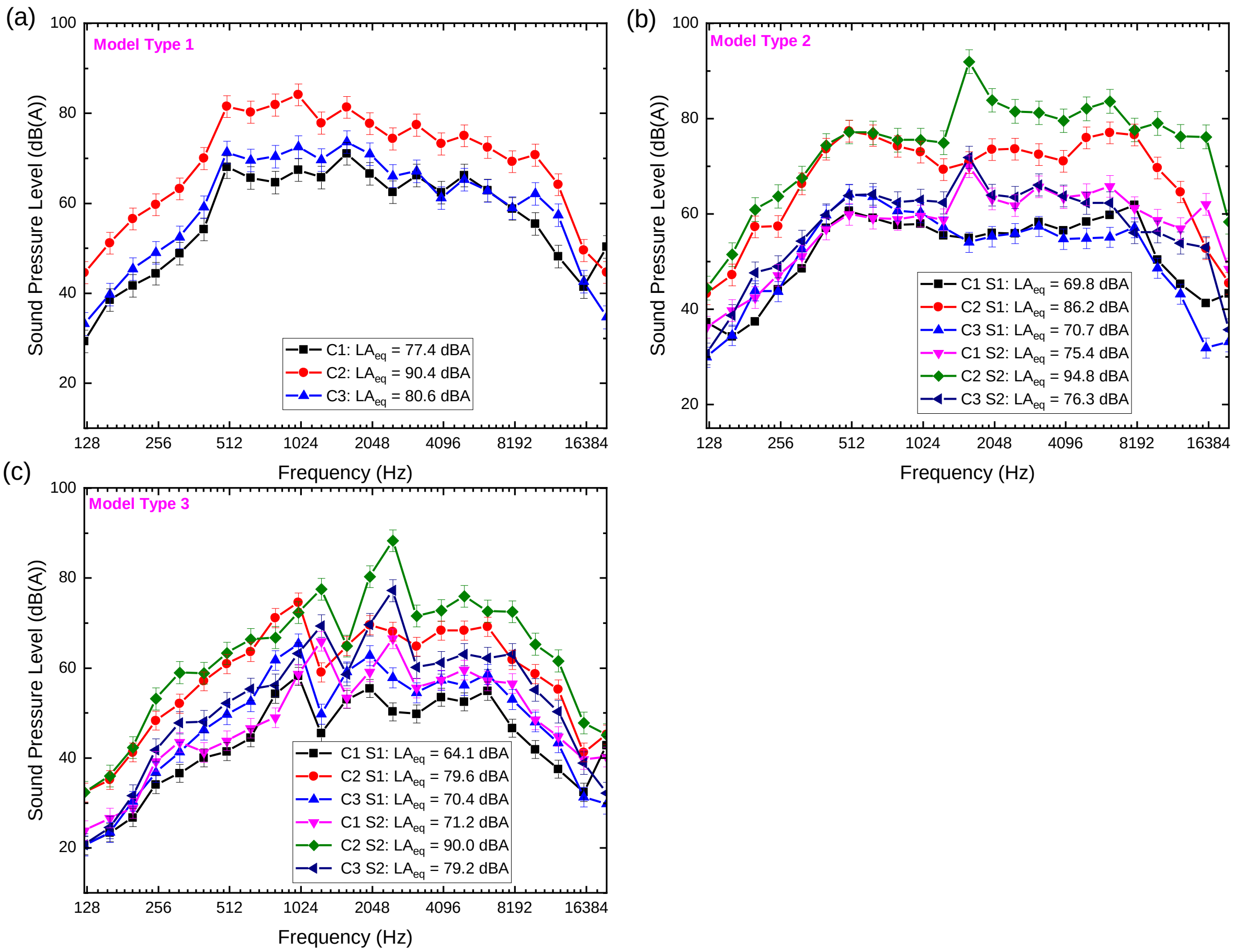

**Figure 2.** Sound pressure level spectra of three different vacuum cleaners (a) Model type 1, (b) Model type 2, (c) Model type 3. All data are plotted on log2 scale. The equivalent sound pressure levels presented in the figure is $LA_{eq(90)}$.

## 4.2 Psychoacoustic performance

**Figure 3** to **Figure 5** shows the results of psychoacoustic metrics (loudness, sharpness, fluctuation strength, and roughness) obtained from the recorded sound levels of vacuum cleaner units. As shown in **Figure 3**(a), the maximum perceived sound intensity is at the user's right ear and is about 109 sone as it is nearest to the vacuum cleaner (model type 1). Moreover, the user's left ear C3 experiences slightly higher loudness than the surrounding occupants C1. Also, C1 has larger sharpness values than C2 and C3, implying the sharpness is the highest for the surrounding occupants (**Figure 3**b). In **Figure 3**(c), the user's right ear (C2) encounters the highest roughness value, in the modulation frequency of 30-300 Hz. Furthermore, the fluctuation strengths captured are nearly identical in all microphones.

The model type 2 surprisingly revealed a maximum loudness value of 118 sones to the user's right ear at the maximum speed setting S2, and a second highest perceived sound intensity was observed to the user's right ear but in minimum rotation speed setting S1 (**Figure 4**a). The other two microphones C1 and C3, have a slight variation in their loudness values in both speed settings. Furthermore, as shown in **Figure 4**(b), at lower speed setting S1, the sharpness value for C3 microphone was lowest over the period, while for the right ear microphone C2, the sharpness value was higher than that for C1 except for few values. A similar trend was observed at higher speed setting S2. Since the user had kept the vacuum cleaners on the right hand during the operation, that may be attributed to the lower sharpness value for the left ear's C3 microphone. Also, there was no statistically significant difference in roughness values between each configuration (**Figure 4**c). The user's right ear microphone attained the slightly higher roughness value of recorded sound at both speed settings. Since the roughness varies with time, it may be possible that the variation with time had a more significant effect than the difference in motor speed and recording position (listening position) compared to other psychoacoustic parameters. Similarly, as shown in **Figure 4**(d), the fluctuation strength values were not significantly dependent on the microphone positions. However, some variations in measured values were observed with the time. Their difference in fluctuation strength was within 0.25 vacil.

For the third vacuum cleaner systems (model type 3), the highest loudness value of 78 sones is recorded in C2 for both speed settings (**Figure 5**a). It can be observed that the sound intensity perceived by the user's right ear C2 is higher than by the left ear C3 and surrounding occupants C1 regardless of the equipment types and speed settings. For sharpness values, the occupant's microphone C1 was perceived to be the highest among all microphones at each setting (**Figure 5**b). While for the right and left ear's microphones, the sharpness values were not significantly different. However, some significant peaks could be observed at several time periods.

Similar to the model type 2 vacuum cleaners, the roughness (**Figure 5**c) and fluctuation strength (**Figure 5**d) were dependent on the operation time for the model type 3 system. These parameters didn't not deviate much between different locations and different speeds, indicating they have close amplitude variations.

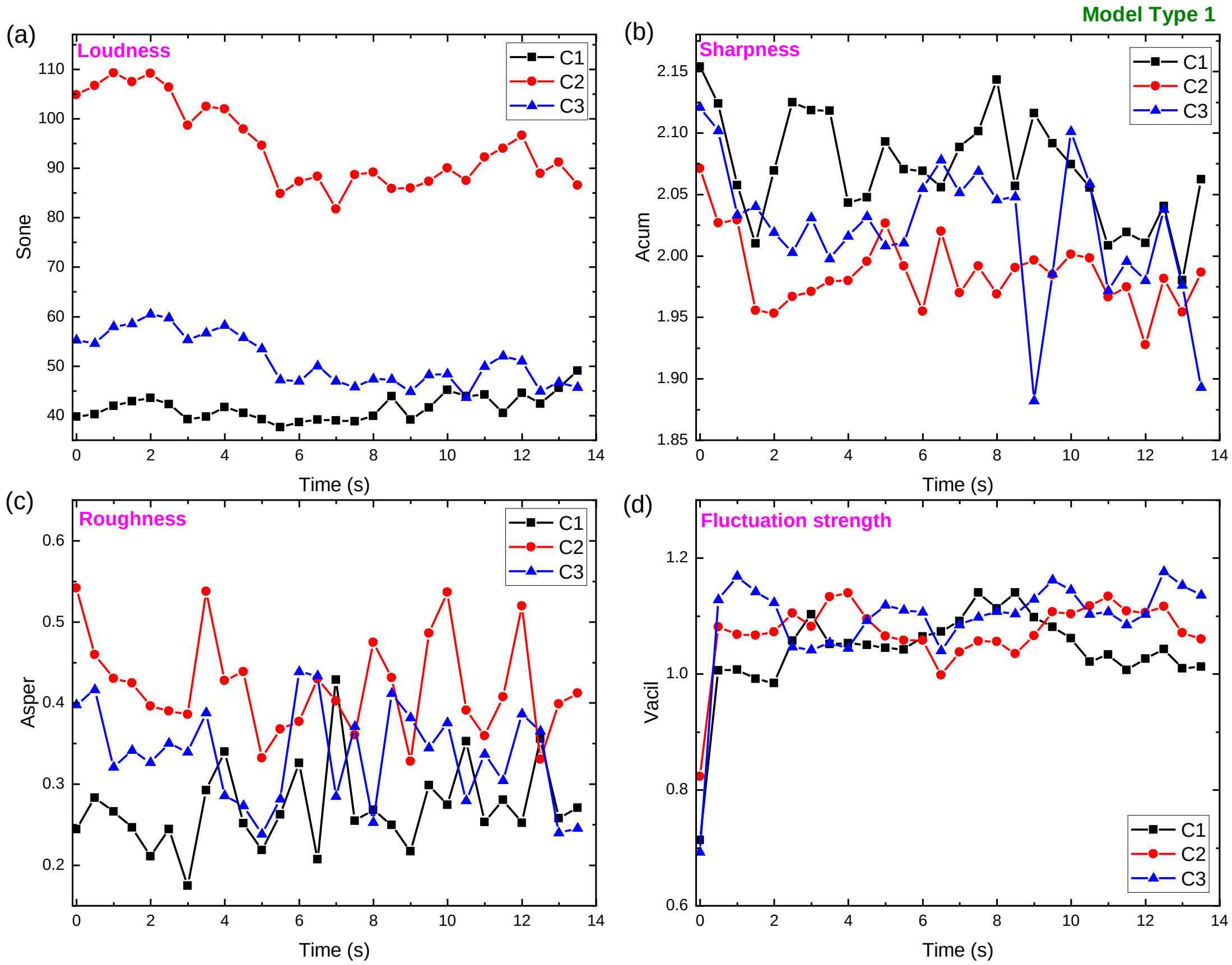

**Figure 3.** Psychoacoustic parametric values for the vacuum cleaner (model type 1). (a) loudness, (b) sharpness, (c) roughness, and (d) fluctuation strength.

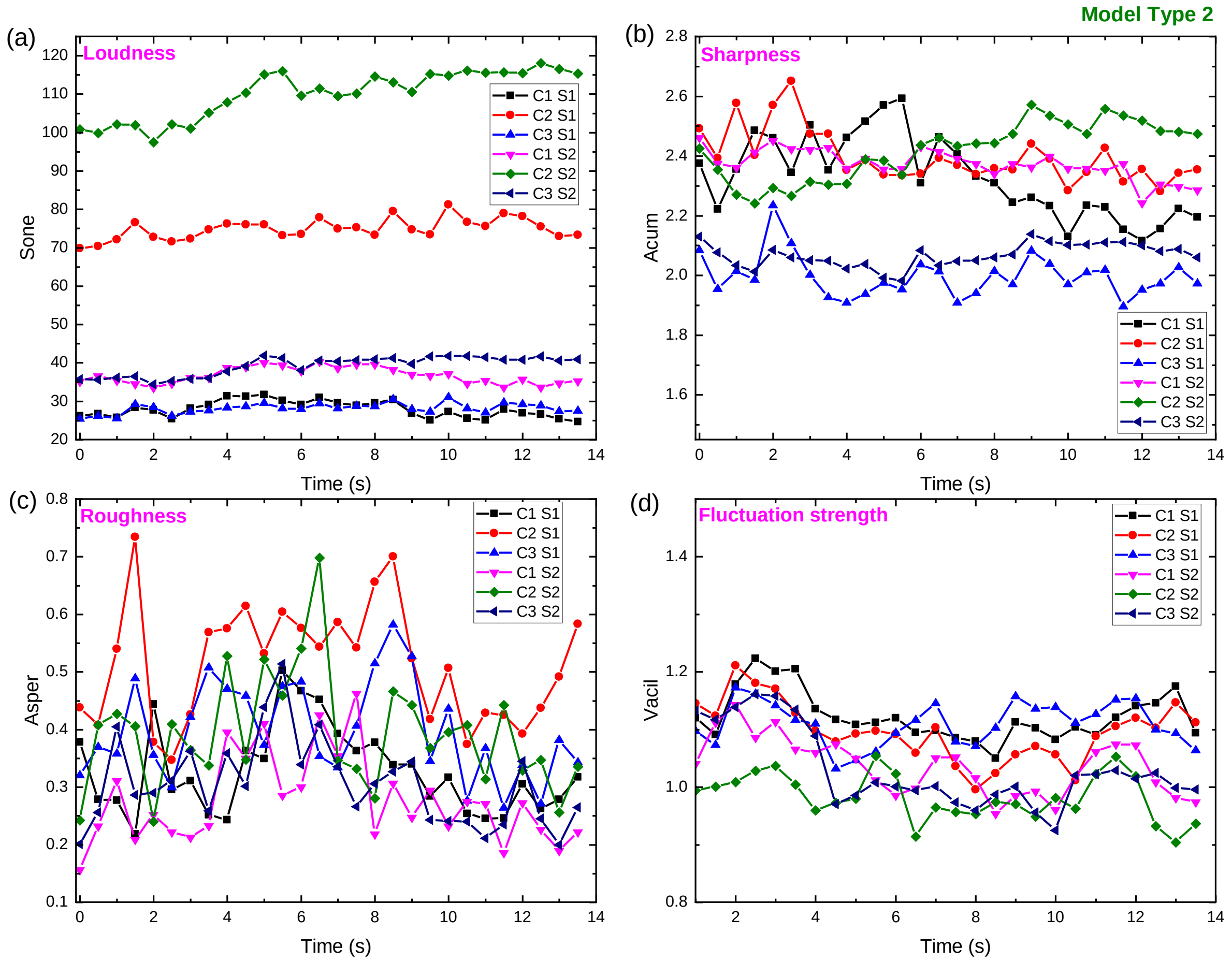

**Figure 4.** Psychoacoustic parametric values for the vacuum cleaner (model type 2). (a) loudness, (b) sharpness, (c) roughness, and (d) fluctuation strength.

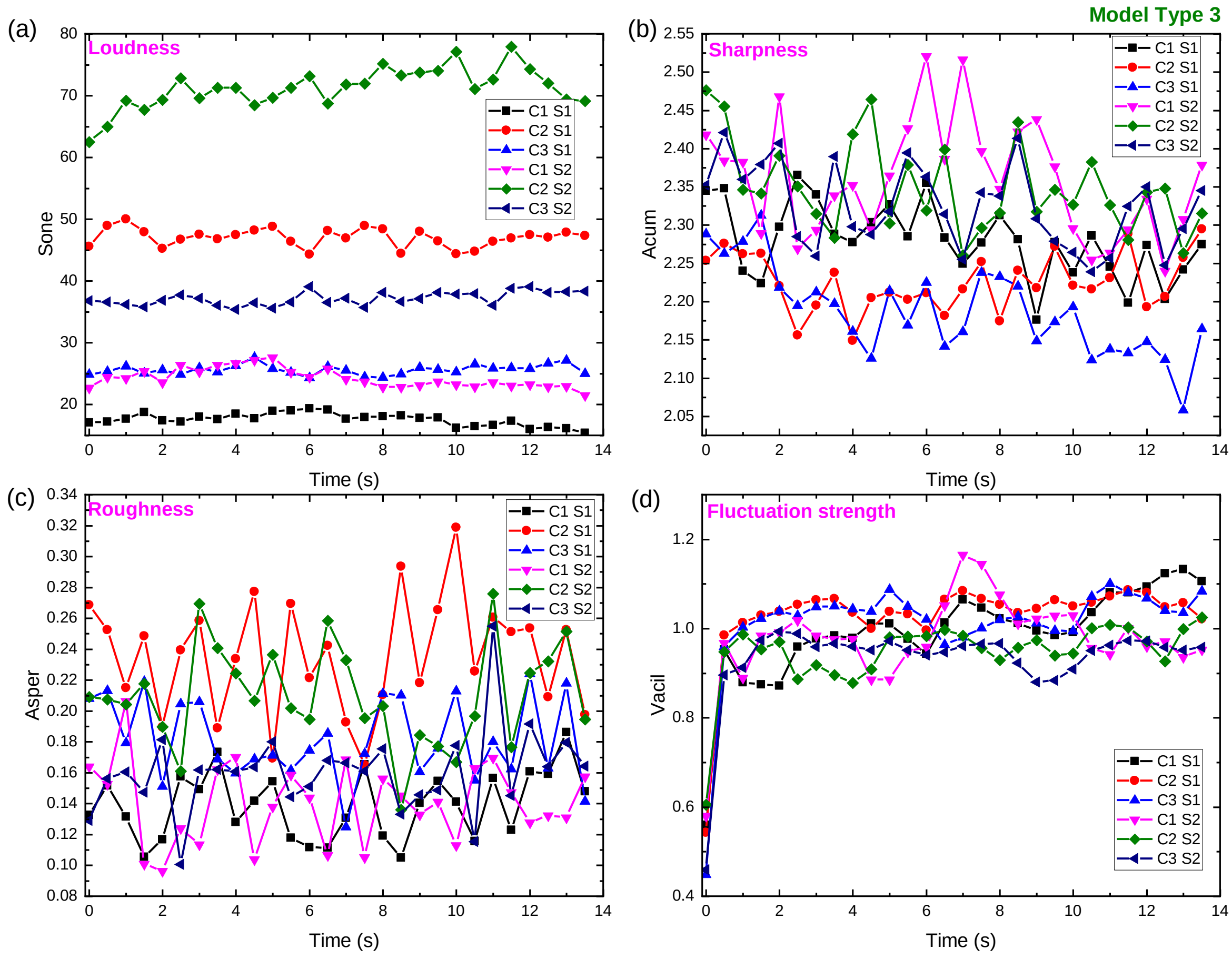

**Figure 5.** Psychoacoustic parametric values for the vacuum cleaner (model type 3). (a) loudness, (b) sharpness, (c) roughness, and (d) fluctuation strength.

## 4.3 Comparative analysis

**Figure 6** presents the comparative results of psychoacoustics parameters for three vacuum cleaners. As shown in **Figure 6**(a), model type 1 produces the highest loudness to the surrounding occupants C1, followed by model type 2 and model type 3 at the maximum and the minimum settings. Besides, the loudness captured at right ear mic C2 is higher than the left ear mic C3. Because the vacuum cleaner was hold on right hand of the user (**Figure 1**a), and the person's body may cause the obstruction into the sound wave path to reach the left ear. The most substantial loudness impact towards the right ear of the user has resulted from model type 2 at the maximum setting (C2S2).

**Figure 6**(b) shows the average sharpness values perceived by surrounding occupants as well as the user's ears. As shown, the average sharpness values perceived by surrounding occupants C1 in the case of model type 1 system was the lowest and model type 2 and model type 3 produced higher sharpness.

Moreover, contrary to loudness, the sharpness measurements did not show a clear differentiation between the left and right ears. The average sharpness values measured by C2 and C3 for model type 1 and model type 3 vacuum cleaners at both settings were approximately the same. However, the average sharpness value in C2 for the model type 2 vacuum cleaner (C2S2) was more significant than that of C3. Right ears received the highest sharpness values for model type 2 vacuum cleaner at both settings.

**Figure 6**(c) shows the average roughness values obtained from the different vacuum cleaners. As shown, roughness values for the model type 3 were almost constant despite different settings across different microphones. The model type 2, however, has shown larger roughness values when operating at the minimum settings across all microphones. Overall, model type 2, at the minimum setting, has the highest roughness values (C1S1 and C2S1). The observations across the left and right ears remained similar across all three devices where the right ear registered the larger roughness values. The fluctuation strength of the vacuum cleaners exhibited small variations across each microphone (**Figure 6**d). The values range from 0.9 to 1.1 vacil for all microphones.

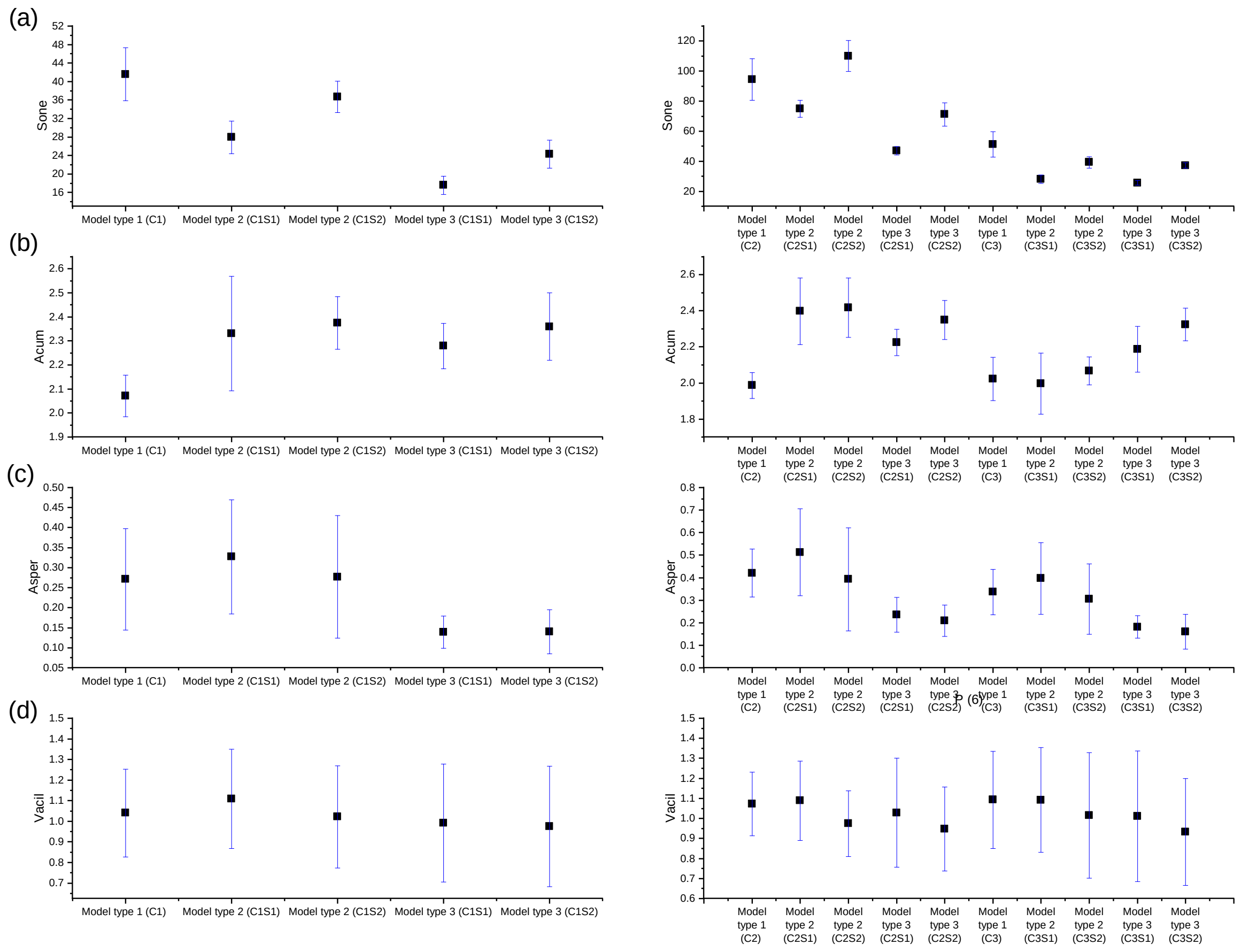

**Figure 6.** Comparative psychoacoustic parametric values for different vacuum cleaners with corresponding standard deviations. (a) loudness (b) sharpness (c) roughness, and (d) fluctuation strength.

## 4.4 Annoyance Index analysis

As described in the aforementioned section 2.4, the relative annoyance index (AI) score for the central, left, and the right microphones are calculated by using the relation proposed by Altmsoy et al. [27]. The central microphone C1 is analogous to individuals present nearby. The C2 and C3 mics are the sensory perceptions of the vacuum noise comparable to the right and left ears of the user. **Figure 7** shows the calculated annoyance index for the recorded sound perceived from these microphones. As shown in **Figure 7**(a), the AI index was found to be highest in the central microphone for the model type 1 system followed by the model type 3. Also, no significant difference in AI index was found for model type 2.

The annoyance index for other microphones for three vacuum cleaners are shown in **Figure 7**(b), model type 2 vacuum cleaners have registered the highest annoyance at its high-speed settings (C2S2). It was expected as vacuum cleaner operated at higher speed settings produced noise with a high sound level. Moreover, the right ear side microphone showed higher annoyance than the left ear microphone for each vacuum cleaner. The possible reason for this trend might be the experimental conditions. During the vacuum cleaning operation, the device was held consistently on the right hand. Hence the right ear was in closer vicinity to the device. As such, this could be the reason for a higher annoyance index values in the right ear. The left and right ear may not perceive the similar quality of sound, and other factors such as the distance of the sound source from each ear can influence the evaluated level of annoyance.

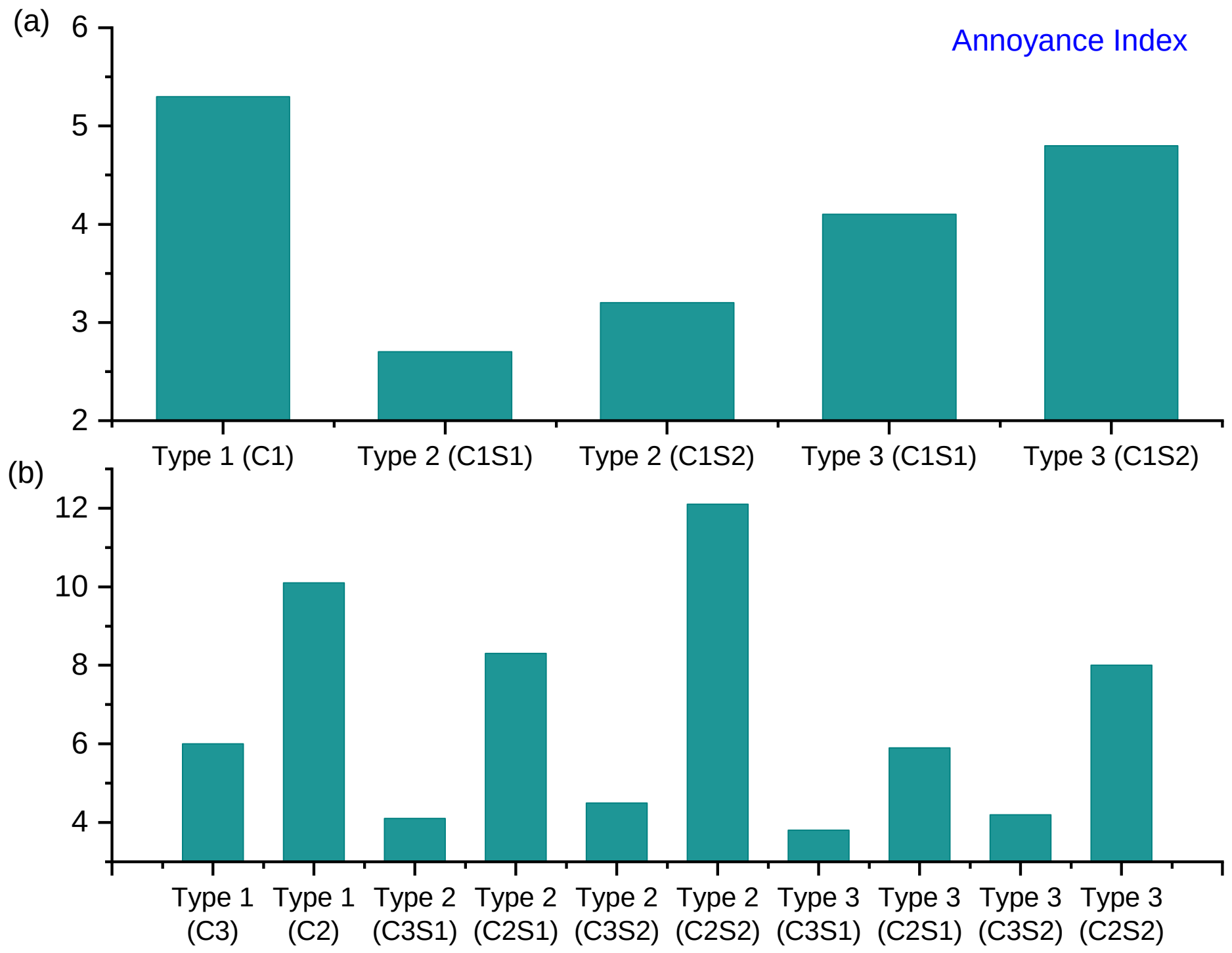

**Figure 7.** Annoyance index for (a) center mic C1, and (b) right mic C2 and left mic C3 for different types of vacuum cleaners. The minimum and maximum motor speed settings are represented by S1 and S2, respectively.

## 5. Conclusions

The presented experimental investigations provide exciting information related to the sound quality index of vacuum cleaners (LG V-CP243NB (model type 1), Dyson Cyclone V10 (model type 2), and Xiaomi Cleanfly Gen 2 (model type 3)). A series of the experiment is carried out to measure the noise produced by each vacuum cleaner. Binaural headphones and a condense microphone are used to emulate the psychoacoustic effect towards the user and the surrounding occupants. Annoyance level index is determined for each vacuum cleaner by using the subjective and objective factors. The analysis of the annoyance level index would result in a better understanding of the human perception of the emitted noise.

Moreover, based on the estimated values, the following observations are made. It is observed that the loudness produced by the device has a more significant effect on its annoyance level, which complies with the previously published works. At the same time, it is also noted that the level of annoyance perceived by the user of the device can be different from those that are within proximity of to the appliance. From the psychoacoustic analysis, the annoyance index of model type 2 at the maximum speed setting is the highest, while the annoyance index of model type 1 is the largest to the surrounding people.

The presented investigations may add a valuable contribution to the assessment of vacuum cleaner noise and may help in the development of a better appliance with the permissible noise level. The studies could be extended to different models of vacuum cleaners and also wet type vacuum cleaners.

## Author contributions

All authors have made valuable contributions to the presented work. S.K. has written the manuscript and helped in experimental planning. W.S.W. and T.M. have conducted the experiments and performed psychoacoustic analysis. H.P.L. conceived the idea and monitored the overall project at every stage.